\documentclass[10pt,twocolumn,letterpaper]{article}

\usepackage{times}
\usepackage{epsfig}
\usepackage{graphicx}
\usepackage{amsmath}
\usepackage{amssymb}

\makeatletter
\newsavebox\myboxA
\newsavebox\myboxB
\newlength\mylenA

\newcommand*\xbar[2][0.75]{%
	\sbox{\myboxA}{$\m@th#2$}%
	\setbox\myboxB\null
	\ht\myboxB=\ht\myboxA%
	\dp\myboxB=\dp\myboxA%
	\wd\myboxB=#1\wd\myboxA
	\sbox\myboxB{$\m@th\overline{\copy\myboxB}$}
	\setlength\mylenA{\the\wd\myboxA}
	\addtolength\mylenA{-\the\wd\myboxB}%
	\ifdim\wd\myboxB<\wd\myboxA%
	\rlap{\hskip 0.5\mylenA\usebox\myboxB}{\usebox\myboxA}%
	\else
	\hskip -0.5\mylenA\rlap{\usebox\myboxA}{\hskip 0.5\mylenA\usebox\myboxB}%
	\fi}
\makeatother

\usepackage{graphics}
\usepackage{algorithmic,algorithm}
\usepackage{amsthm}                 
\usepackage[mathscr]{eucal}         
\usepackage{amsbsy}                 
\usepackage{bm}                     
\usepackage{paralist}               
\usepackage{xspace}                 

\newcommand{\barr}{\left[ \begin{array} }
\newcommand{\earr}{ \end{array} \right] }
\newcommand{\ars}[1]{\left[ \begin{array}{#1}}
\newcommand{\are}{\end{array} \right] }
\newcommand{\oars}[1]{\begin{array}{#1}}
\newcommand{\oare}{\end{array}}
\newcommand{\eqs}{\begin{eqnarray}}
\newcommand{\eqe}{\end{eqnarray}}
\newcommand{\eqsn}{\begin{eqnarray*}}
\newcommand{\eqen}{\end{eqnarray*}}

\newcommand{\ens}{\begin{enumerate}}
\newcommand{\ene}{\end{enumerate}}
\newcommand{\its}{\begin{itemize}}
\newcommand{\ite}{\end{itemize}}
\newcommand{\des}{\begin{description}}
\newcommand{\dee}{\end{description}}

\numberwithin{theorem}{subsection}

\numberwithin{lemma}{subsection}

\numberwithin{defn}{subsection}

\numberwithin{conjecture}{subsection}

\numberwithin{remark}{subsection}

\begin{document}

\title{Effects of Home Resources and School Environment on Eighth-Grade Mathematics Achievement in Taiwan }

\author{Jiaqi Cai\\
Boston College\\
}

\maketitle
\thispagestyle{empty}

\section{Problem Statement}
Over the past decades, researchers have explored the relationship among home resources, school environment, and students' mathematics achievement in a large amount of studies (Coleman, 1966; Greenwald, Hedges \& Laine, 1996; Hampden-Thompson \& Johnston, 2006; Mullis et al., 2012). Many of them suggested that rich home resources for learning were related to higher average academic achievement. Some also suggested that the home background was closely associated with the learning environment, and therefore, influenced students' achievements. Thus, this study hypothesized that students who own more home resources would perform better than students who possess fewer resources and that schools that have more socioeconomically advantaged students, located in high-income neighborhoods, and possess more instructional resources would have better mathematics performance. 
The study focuses on eighth graders in Taiwan and explores the variance in mathematics achievement of students as a function of student and school-level differences. It attempts to answer three research questions: 1) How much of the total variance in mathematics achievement of eighth graders in Taiwan have been accounted for student and school-level variance? 2) How much of the student-level variance in mathematics achievement of eighth graders in Taiwan was associated with parents' education level and home educational possessions? 3) How much of the school-level variance in mathematics achievement of eighth graders in Taiwan was associated with students' economic background of school, school location, and school instructional resources? The results, hopefully, will provide some means to help parents, educators, and policy makers in Taiwan improve students' math achievement and narrow the achievement gap among schools.

\section{Literature Review}
The home resource has been considered a key factor in students' academic performance. Coleman's report (1966) presented compelling evidence of the pivotal role the home resource plays in children's cognitive skill development. Later, a series of research studies were conducted and supported the importance of home resource in student learning outcome. For instance, Hampden-Thompson \& Johnston (2006) found that parents' education level and occupational status, and the number of books at home were positively related to students' academic achievement. Mullis et al. (2012) reported a similar finding on the Trends in International Mathematics and Science Study (TIMSS) 2011 International Results in Mathematics as well: parents' education level and home educational possessions, such as books, study rooms, and Internet connections were the important resources that were highly related to student achievement. Therefore, this study is intended to examine how much variance in math achievement of students of eighth grade in Taiwan was associated with students' home resources. 
Apart from student-level factors, school resources were also proven to be highly correlated with students' academic achievement. Research of Greenwald, Hedges, \& Laine (1996) confirmed that ``school resources are systematically related to student achievement and that these relations are large enough to be educationally important'' (p.384). Moreover, Mullis et al. (2012) found that successful schools tended to have students that were relatively economically affluent, possessed more instructional materials, and located in a better location. This study, therefore, included several school-level variables that were key elements of school resources to investigate how much variance in math achievement could be explained by students' economic background of school, school location, and math instructional resources.

\section{Methods}
\subsection{Data and Samples}
TIMSS \& PIRLS International Study Center is a research center at Boston College that conducts a series of assessments in a number of countries to measure trends in mathematics and science achievement at the fourth and eighth grades. In general, TIMSS assessments include achievement tests as well as questionnaires for students, parents, teachers, schools, and curricula. 
This study used TIMSS 2011 student achievement data and selected a few items from student and school questionnaires. The data was used to investigate how student-level factors and school-level factors affect math achievement of eighth graders in Taiwan. The data was gathered from the responses from 5,042 students of eighth grade who attended 150 schools in Taiwan, as well as the responses from 150 school principals. The responses from 4,605 students of 140 schools were eventually included in the current study after handling missing data.

\subsection{Variables}
\subsubsection{Student Level Measures}
As the previous section mentioned, research consistently showed a strong positive relationship between home resources and academic achievement. According to the TIMSS 2011 report, parents' education background and home educational possessions, such as books, Internet connections, and study rooms, were key components of home resources. Therefore, mother's education background, father's education background, and home educational possessions are three student-level factors in this analysis. \\

\emph{Mathematics Achievement (mat)}. The math ability of students of eighth grade in Taiwan was measured in four content areas; they were number, algebra, geometry, data, and chance. Five plausible values provided by TIMSS 2011 were used in this study. \\

\emph{Mother's Education Background (mo)} \& \emph{Father's Education Background (fa)}. Both mother and father's education background were acquired in the student questionnaire. Students were instructed to pick one option that applied to his or her parents' highest level of education. In order to specify the influence from mother and father, mother's education background and father's education background were two variables measured at the individual level in this study. Both variables were recoded for the analysis by following the rules:

\begin{center}
	\begin{tabular}{| l | c |}
		\hline
		Highest Education Level of Parents & Points\\ \hline
		Bachelor's Degree or higher & 5  \\ \hline
		Associate's Degree & 4  \\ \hline
		High school & 3  \\ \hline
		Lower-Secondary & 2  \\ \hline
		Primary or lower& 1 \\ \hline
	    The student doesn't know & 0 \\ \hline
	\end{tabular}
\end{center}

\emph{Home Possessions (hp)}. Home Possessions is a composite variable in the current analysis. It consists of six survey items, which ask if the participant has a computer, a study desk, his or her own books or room, Internet connections, and DVDs, software, and videos for mathematics and science learning at home. Each item was recoded to: 1 point if the answer was 'Yes'; 0 point if the answer was 'No'. The value of home possessions variable was the sum of six questionnaire items' scores. Thus, for each participant, the higher score he or she obtained in this variable, the more home educational possessions the participant had at home. 

\subsubsection{School Level Measures}
\emph{Student Economic Background (stueco)}. Student Economic Background is a composite variable, which contains two school questionnaire items: (1) how many percentages of students come from economically disadvantaged homes, and (2) how many percentages of students come from economically affluent homes. The variable was recoded for analysis by applying the following rules:  

\begin{center}
	\begin{tabular}{| l | c |}
		\hline
		Student Economic Background of School & Points\\ \hline
		More affluent than disadvantaged students  & 1  \\ \hline
		More disadvantaged than affluent students & -1  \\ \hline
    	Neither case & 0  \\ \hline
	\end{tabular}
\end{center}

\emph{School Location (schlo)}. This variable measured the average income level of the school's immediate area. The variable was recoded to: 1 point if the school located in high income level area; 0 point if the school located in medium income level area; and -1 point if the school loacted in low income level area. \\

\emph{School Resources in Computer (schrc)}. A school principal completed the survey item by providing the number of computers that the school provided to eighth graders for math instructional purposes. 

\section{Analysis and Results}
\subsection{Descriptive Analyses}
The descriptive statistics of all variables were presented on Table 1. Students' math achievement score is 612.5 point on average. The mean mother's highest education level is very similar to father's; both are at the  Lower-Secondary level. Students on average gain a value of 4.53 in home possessions variable, indicating that eighth graders in Taiwan own more than half of the resources mentioned in the six items. For variables at the school-level, the mean of Student Economic Background is .04, meaning that the schools in the sample on average have neither more affluent nor more disadvantaged students. Moreover, the mean for School Location is - .09, which indicates that schools on average were located in a medium to slightly low-income neighborhood. Last, eighth graders of each school in the sample own 66.24 computers on average for math instructional purposes. \\

The correlations between different variables in this study are displayed on Table 2. All correlations are significant at the 0.01 level. 

\subsection{Preliminary Analyses}
In this analysis, a two-level Means-as-Outcomes regression model was used to model the relationship between outcome variable math achievements, three student-level predictors (mother's education background, father's education background, and home possessions), and three school-level predictors (student economic background, school location, and school resources in computer).\\
   
Prior to building the most appropriate model to predict students' math achievement, the unconditional model (Model 0) was run in order to figure out what percent of the variability exists between individuals and what percent exists between schools. This was done in order to see if there is nesting within groups in the dataset. Based on Table 3, the reliability of the intercept in this model is very high (0.898). Table 4 shows that the intercept of the unconditional model is 609.14, which is the predicted value of grand mean math achievement score across students and schools. It is statistically significantly different from zero $(p<0.001)$. Also, the variance components are 8195.39 at the student-level and 2238.60 at the school-level. Both of the random effects are statistically significant $(p<0.001)$.  \\

\noindent \emph{Unconditional Model (Model 0)}\\

\noindent Level-1 Model:
\begin{equation}
\nonumber
Y_{ij} = \beta_{0j} + \gamma_{ij}
\end{equation}
\\
Level-2 Model:
\begin{equation}
\nonumber
\beta_{0j} = \gamma_{00} + \mu_{0j}
\end{equation}
\\
The mixed model:
\begin{equation}
\nonumber
\begin{aligned}
Y_{ij} = &\gamma_{00} + \mu_{0j} + \gamma_{ij} \\
Y_{ij} = &609.14+\mu_{0j}+\gamma_{ij}
\end{aligned}
\end{equation}

In order to check if there is nesting that needs to be taken into account, the unconditional ICC was calculated based on the Table 5 results. This tells us the proportion of variance that exists at the school-level. If this number is very low, an OLS regression method will be appropriate instead of a multilevel modeling because there is no nesting to take into account. 

\begin{equation}
\nonumber
\text{ICC} = \frac{\widehat{\tau}_{00}}{\widehat{\tau}_{00}+\widehat{\sigma}^2} = \frac{2238.6}{2238.6+8195.38} = 0.215
\end{equation}

\noindent This means that $21.5\%$ of the variance in mathematics achievement of eight graders in Taiwan lied between schools. This nesting was tested with a chi-square statistics, which was significant $(p<0.001)$, indicating that there was a significant amount of variability to be explained between schools. Therefore, a multilevel model will be constructed. The school-level variables in the model will attempt to explain some of this variance that exists between schools. The individual-level variables will attempt to explain the $78.5\%$ of the variance that exists between individuals. 

Next, in order to assess the appropriate statistical power, the effective sample size was calculated. The design effect was computed as follows,
\begin{equation}
\nonumber
1+(n-1)\times \text{ICC} = 1+(32.89-1) \times 0.215 = 7.86
\end{equation}

Then we got,
\begin{equation}
\nonumber
\text{Effective Sample Size} = \frac{N}{\text{Design Effect}} = \frac{4605}{7.86} = 586
\end{equation}
Therefore, the effective sample size is 587, which indicates a sufficient power for these analyses.

\subsubsection{Level-1 Analyses}
This section focuses on developing and finalizing the level-1 model for the sample. Mother's education background \emph{(mo)}, father's education background \emph{(fa)}, and home possessions \emph{(hp)} are the level-1 predictors that used to predict students' math achievement. In the Model 1, \emph{mo} and \emph{fa} were added using grand-mean centering. All slopes were not allowed to vary randomly across schools. The reliability of the predicted intercept is high (0.877). The results of this model indicate that both \emph{mo} $(\gamma = 6.86, SE = 1.19, p<0.001)$ and \emph{fa} $(\gamma = 8.19, SE=1.10, p<0.001)$ are significant predictors to students' mathematics performance. The slope of \emph{mo} means that for one-unit increase in mother's education background above the grand mean, partialling out father's education background, there is a predicted 6.86 point increase in math achievement score. Similarly, the regression coefficient of \emph{fa} is 8.19, indicating that when controlling for mother's education background, one-unit increase in father's education background above the grand mean leads to a predicted 8.19 point increase in student's math achievement. The variance components indicate that $4\%$ of the variance at the individual-level and therefore $3\%$ of the total variance in math achievement were explained by mother's education background and father's education background.

The third level-1 predictor, home possessions \emph{(hp)}, was added to the model. The reliability of this new model (Model 2) is also high (0.848). The results of this model indicate that \emph{hp} $(\gamma = 9.29, SE=1.41, p<0.001)$ is a significant predictor in the model. The coefficient means that when controlling for \emph{mo} and \emph{fa}, one-unit increase in home possessions index above the grand mean leads to a predicted 9.29 point increase in student's math score. Also, predictors \emph{mo} $(\gamma = 6.24, SE = 1.17, p<0.001)$ and \emph{fa} $(\gamma = 7.45, SE=1.07, p<0.001)$ remain significant in this model. Therefore, we can conclude that \emph{hp}, \emph{mo}, and \emph{fa} are significant student-level predictors in the sample. Adding the home possessions variable also helped explain another $1\%$ of the individual-level variance; so this model together explained $5\%$ of the variance at the individual-level and $4\%$ of the total variance in math achievement. 

In the Model 3, the slopes of the three level-1 predictors were allowed to vary randomly across schools in order to see if there is any variability in slopes between schools that can be predicted at Level-2. The results of this model show that \emph{hp}, \emph{mo}, and \emph{fa} significantly predict math achievement, but their slopes do not vary between schools $(p>0.05)$. This indicates that there is no variability in slopes of the level-1 predictors among schools. Therefore, the slopes of these three variables are fixed and the finalized level-1 model is shown below.\\

\noindent \emph{Model 4}\\

\noindent Level-1 model:
\begin{equation}
\nonumber
\begin{aligned}
Y_{ij} = &\beta_{0j} + \beta_{1j}(\text{mo}_{ij} - \xbar{\text{mo}}.. ) + \beta_{2j}(\text{fa}_{ij}-\xbar{fa}..) + \\
&\beta_{3j}(\text{hp}_{ij}-\xbar{hp}..) + \gamma_{ij}
\end{aligned}
\end{equation}

\noindent Level-2 model:
\begin{equation}
\nonumber
\begin{aligned}
\beta_{0j} &= \gamma_{00} + \mu_{0j} \\
\beta_{1j} &= \gamma_{10} \\
\beta_{2j} &= \gamma_{20} \\
\beta_{3j} &= \gamma_{30}
\end{aligned}
\end{equation}

\noindent The mixed model:
\begin{equation}
\nonumber
\begin{aligned}
Y_{ij} = &\gamma_{00} + \gamma_{10}(\text{mo}_{ij} - \xbar{\text{mo}}.. ) + \gamma_{20}(\text{fa}_{ij}-\xbar{fa}..) + \\
&\gamma_{30}(\text{hp}_{ij}-\xbar{hp}..) + \mu_{0j} + \gamma_{ij} \\
Y_{ij} = &609.72 + 6.24(\text{mo}_{ij} - \xbar{\text{mo}}.. ) + 7.45(\text{fa}_{ij}-\xbar{fa}..) + \\
&9.29 (\text{hp}_{ij}-\xbar{hp}..) + \mu_{0j} + \gamma_{ij} \\
\end{aligned}
\end{equation}

\subsubsection{Level-2 Analyses}
After finalizing the level-1 model, the school-level predictors were added for the analyses. The level-2 predictors used in this model (Model 5) were student economic background \emph{(stueco)}, school location \emph{(schlo)}, and school resources in computer \emph{(schrc)}. Since there was no evidence from previous studies that the level-2 predictors were associated with mother's education background, father's education background, and home possessions, the level-2 variables were only entered to predict the intercept. The results of the Model 5 show that along with the student-level predictors, \emph{stueco} $(\gamma = 19.41, SE = 6.54, p=0.004)$, \emph{schlo} $(\gamma = 23.41, SE = 9.31, p=0.013)$, and \emph{schrc} $(\gamma = 0.2, SE = 0.08, p=0.016)$ are significant predictors between school differences in math achievement (Table 9). 

Based on the above analyses, Model 5 is the final model for the sample. In this model, the reliability of the intercept is 0.836, which is desirable. The intercept is equal to 597.93, meaning that if a student whose mother's highest education level and father's highest education level equals the grand means, home possessions is identical to the grand mean, and he or she is in a school that has neither more affluent nor more disadvantaged students, locates in medium income level area, and owns zero computer for math instructional purposes by eighth-grade students, the predicted math score of this student is 597.93 points. The results of this model also indicate that mother's education background $(\gamma = 6.21, SE = 1.17, p = <0.001)$, father's education background $(\gamma = 7.32, SE = 1.07, p = <0.001)$, home possessions $(\gamma = 9.09, SE = 1.40, p = <.001)$, student economic background $(\gamma = 19.41, SE = 6.54, p = 0.004)$, school location $(\gamma = 23.41, SE = 9.31, p = .013)$, and school resources in computer $(\gamma = 0.2, SE = 0.08, p = 0.016)$ are significant predictors of math achievement (Table 9). The coefficient for student economic background indicated that for every one-unit increase in school student economic background above the grand mean, math achievement score is predicted to increase by 19.41 points holding all other variables constant. The coefficient for school location means that for every one-unit increase in school location index, there is a predicted 23.41 points increase in math achievement score holding all other variables constant. Last, the slope of school resources in computer indicate that by holding all other variables constant, every one additional computer owned by eighth-grade students in the school for math instructional purposes leads to a predicted 0.2 point increase in math achievement score. Also, this final model explained 5\% of the student-level variance, 44.8\% of the school-level variance, and 13.6\% of the total variance (Table 8). There was still a significant amount of variance at both levels left to be explained. 

The final model is presented below: \\

\noindent Level-1 model:
\begin{equation}
\nonumber
\begin{aligned}
Y_{ij} = &\beta_{0j} + \beta_{1j}(\text{mo}_{ij} - \xbar{\text{mo}}.. ) + \beta_{2j}(\text{fa}_{ij}-\xbar{fa}..) + \\
&\beta_{3j}(\text{hp}_{ij}-\xbar{hp}..) + \gamma_{ij}
\end{aligned}
\end{equation}

\noindent Level-2 model:
\begin{equation}
\nonumber
\begin{aligned}
\beta_{0j} = \gamma_{00} + \gamma_{01}W_{j\hspace{1mm}stueco} + &\gamma_{02}W_{j \hspace{1mm} schlo} + \gamma_{03}W_{j\hspace{1mm}schrc} + \mu_{0j}\\
\beta_{1j} &= \gamma_{10} \\
\beta_{2j} &= \gamma_{20} \\
\beta_{3j} &= \gamma_{30}
\end{aligned}
\end{equation}

\noindent The mixed model:
\begin{equation}
\nonumber
\begin{aligned}
Y_{ij} = &\gamma_{00} + \gamma_{01}W_{j \hspace{1mm} stueco} + \gamma_{02}W_{j \hspace{1mm} schlo} + \gamma_{03}W_{j \hspace{1mm} schrc} \\
& +\gamma_{10}(\text{mo}_{ij} - \xbar{\text{mo}}.. ) + \gamma_{20}(\text{fa}_{ij}-\xbar{fa}..) \\
&  + \gamma_{30}(\text{hp}_{ij}-\xbar{hp}..)+ \mu_{0j} + \gamma_{ij} \\
Y_{ij} = &597.93 + 19.41 W_{j \hspace{1mm} stueco} + 23.41W_{j \hspace{1mm} schlo} + 0.2 W_{j \hspace{1mm} schrc}\\
& + 6.21(\text{mo}_{ij} - \xbar{\text{mo}}.. ) + 7.32(\text{fa}_{ij}-\xbar{fa}..)  \\
& + 9.09(\text{hp}_{ij}-\xbar{hp}..)  + \mu_{0j} + \gamma_{ij} \\
\end{aligned}
\end{equation}

\section{Conclusion}

In general, these results indicate that all student-level variables and school-level variables are significant predictors in this study and positively related to mathematics achievement, which are identical to previous research findings. The higher the student's mother or father's highest level of education, the better the student is predicted to perform in math. Another student-level variable, home possessions, is a significant predictor of math achievement as well. The more home possessions a student owns, the higher score the student is expected to achieve in math. Last, there is no variability in slopes of the level-1 predictors between schools. 

In Taiwan, schools with more students from affluent families are predicted to outperform those with fewer numbers of socioeconomically advantaged students. Schools in high income level neighborhoods are expected to perform better in mathematics than those in poor areas. The very small coefficient of school resources in computer indicates that this variable can only predict a very little amount of math achievement, which is not consistent with previous findings. Also, from the findings we can observe that in Taiwan, wide achievement gaps in mathematics exist among schools with different percentages of economically advantaged students. Achievement gaps are also found among schools located in different income level neighborhoods. 

In general, the results imply that educators and policy makers have several possible means to help students of eighth grade in Taiwan improve their math achievement. For instance, they can encourage parents to purchase more educational resources for children and improve the home learning environment. Furthermore, some issues are recommended for future research, for example, examining how parents' education background affect students' math achievement and the performance in other subjects. By understanding this issue, appropriate supports or interventions can be implemented and better assist students. Further investigating the cause of wide math achievement gaps among schools (e.g., funding systems, teacher qualifications, and school organizations) are also needed. If a lack of resources is the cause, administrators or policy makers may consider providing more funding to the schools who need more supports. 

There is a limitation to be noted regarding this study. Extensive research has shown that school instructional resources are important positive indicators for academic achievement. However, school resources in computer only predict a small amount of math achievement in the current study. After examining the content of the related questionnaire item, I suggest including more information into this item in the future. Except for the number of computers owned by eighth graders for math instructional purposes, other instructional resources in schools can be taken into account, for example, math learning tools and software. By including more relevant information, a better understanding of the relationship between school resources for math instruction and math achievement will be possibly obtained. 

Moreover, the method used in this paper could be extended to multi-dimensional data based on techniques such as CP decomposition, Tucker decomposition and t-SVD, for example. More details are under preparation and would be in our next paper.

\section*{Reference}
\nonumber
\noindent Coleman, J.S., Kelly, D.L., Hobson, C.J., McPartland, J., Mood, A.M., Weinfeld, F.D., and     
York, R.L. (1966). Equality of Educational Opportunity. U.S. Department of Health, 
Education, and Welfare. Washington, DC: U.S. Government Printing Office. \\

\noindent Entwisle, D., \& Alexander, K. (1992). Summer Setback: Race, Poverty, School Composition, 
and Mathematics Achievement in the First Two Years of School. American Sociological 
Review, 57(1), 72-84. Retrieved from 
http://www.jstor.org.proxy.bc.edu/stable/2096145.\\

\noindent Greenwald, R., Hedges, L., \& Laine, R. (1996). The Effect of School Resources on Student 
Achievement. Review of Educational Research, 66(3), 361-396. Retrieved from 
http://www.jstor.org.proxy.bc.edu/stable/1170528
Greenwald, R., Hedges, L.V., \& Laine, R.D. (1996). The effect of school resources on student 
achievement. Review of Educational Research, Vol. 66, No. 3 (Autumn, 1996), pp. 
361-396.\\

\noindent Zhang, Z., Aeron, S. (2017). Exact tensor completion using t-svd. \\

\noindent Hampden-Thompson, G., \& Johnston, J. (2006). Variation in the relationship between non- 
school factors and student achievement on International assessments. Statistics in brief. 
Washington, D.C.: U.S. Department of Education, Institute of Education Sciences. 
Retrieved from: http://nces.ed.gov/pubs2006/2006014.pdf. \\

\noindent Zhang, Z., Ely, G., Aeron, S., Hao, N., Kilmer, M. (2014). Novel methods for multilinear data completion and de-noising based on tensor-SVD. \\

\noindent Ma, X., \& Klinger, D. (2000). Hierarchical Linear Modelling of Student and School Effects on 
Academic Achievement. Canadian Journal of Education / Revue Canadienne De 
L'éducation, 25(1), 41-55. doi:1. Retrieved from 
http://www.jstor.org.proxy.bc.edu/stable/1585867 doi:1.\\

\noindent Mullis et al. (2012). TIMSS 2011 International Results in Mathematics. Chestnut Hill, MA: 
TIMSS \& PIRLS International Study Center, Boston College.

\onecolumn
\section*{Appendix}

\begin{figure}[htbp]
	\centering
	\includegraphics[width=\textwidth]{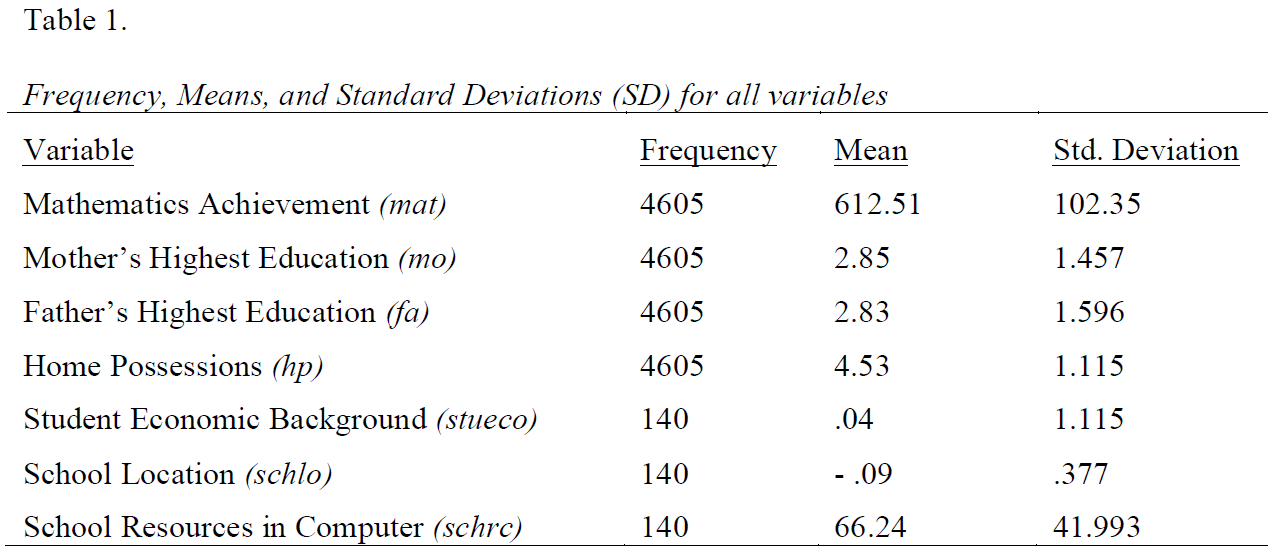}
\end{figure}

\begin{figure}[htbp]
	\centering
	\includegraphics[width=\textwidth]{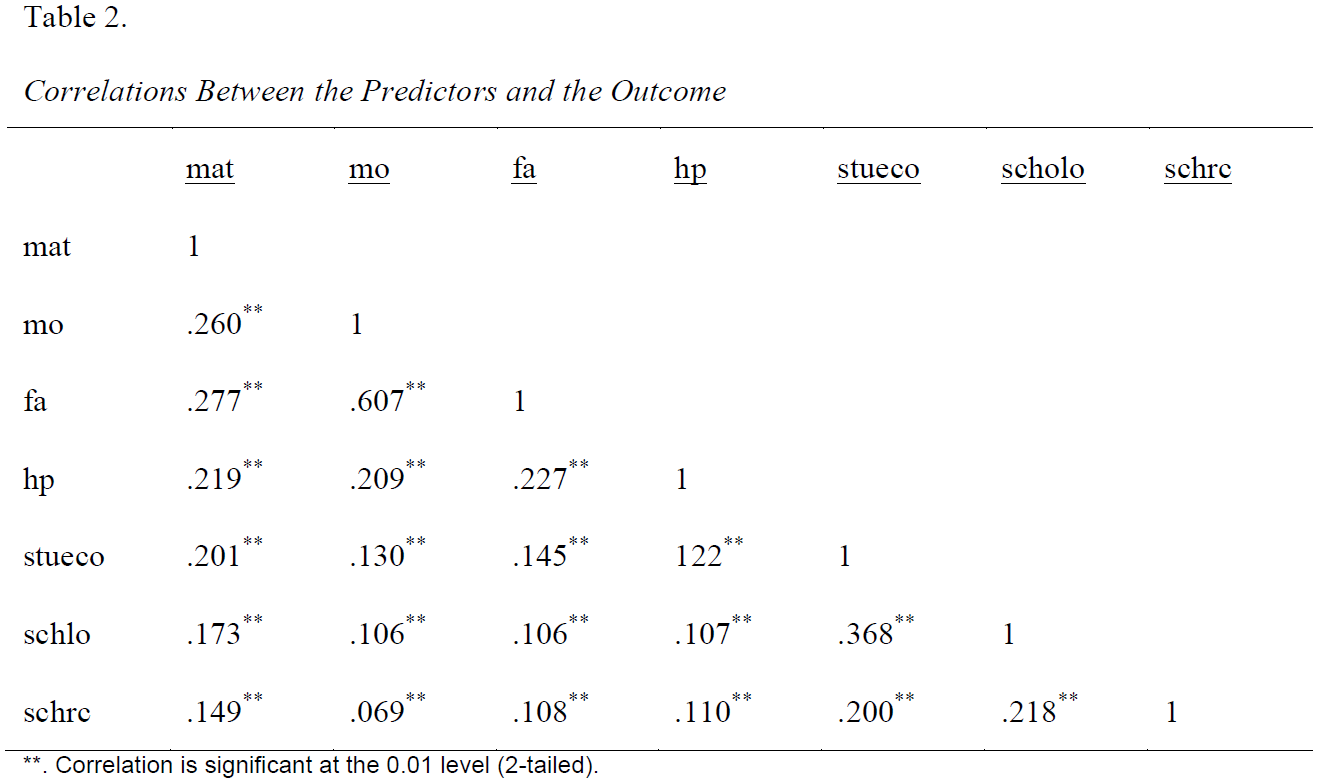}
\end{figure}

\begin{figure}[htbp]
	\centering
	\includegraphics[width=\textwidth]{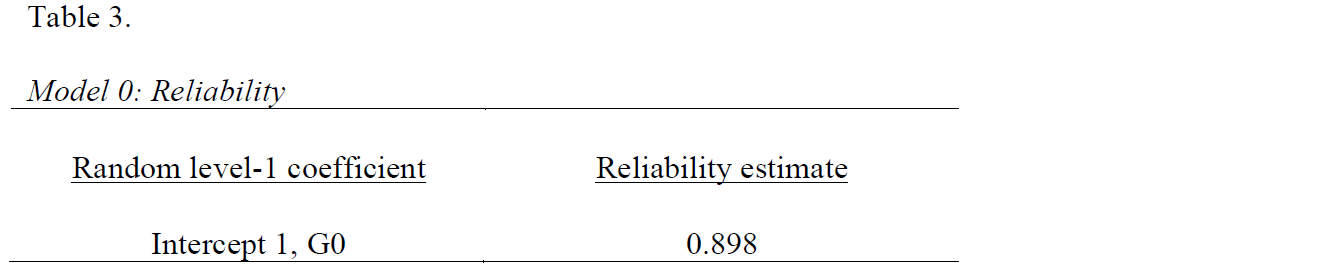}
\end{figure}

\begin{figure}[htbp]
	\centering
	\includegraphics[width=\textwidth]{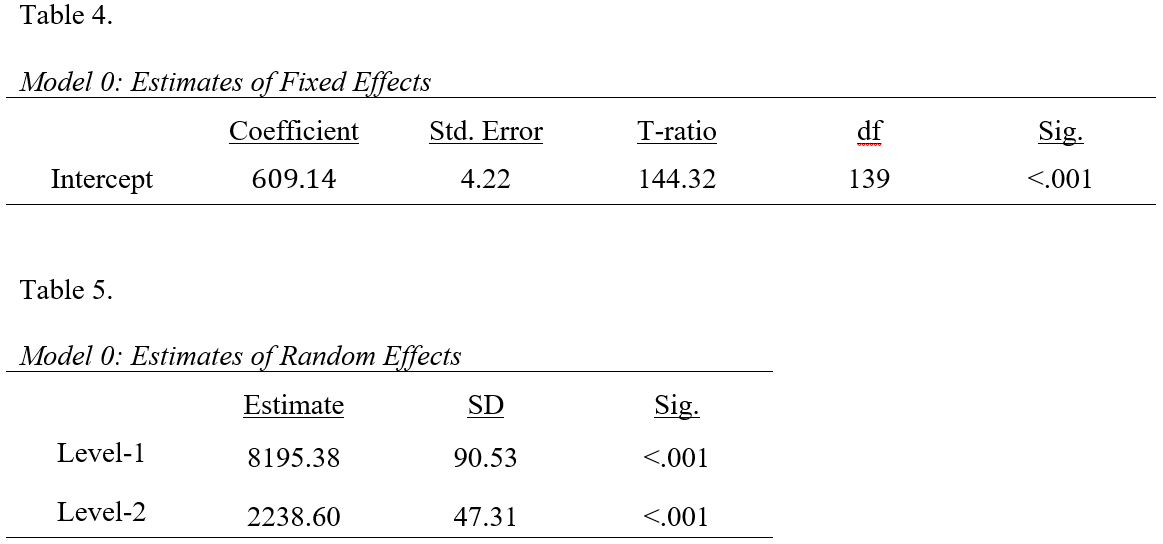}
\end{figure}

\begin{figure}[htbp]
	\centering
	\includegraphics[width=\textwidth]{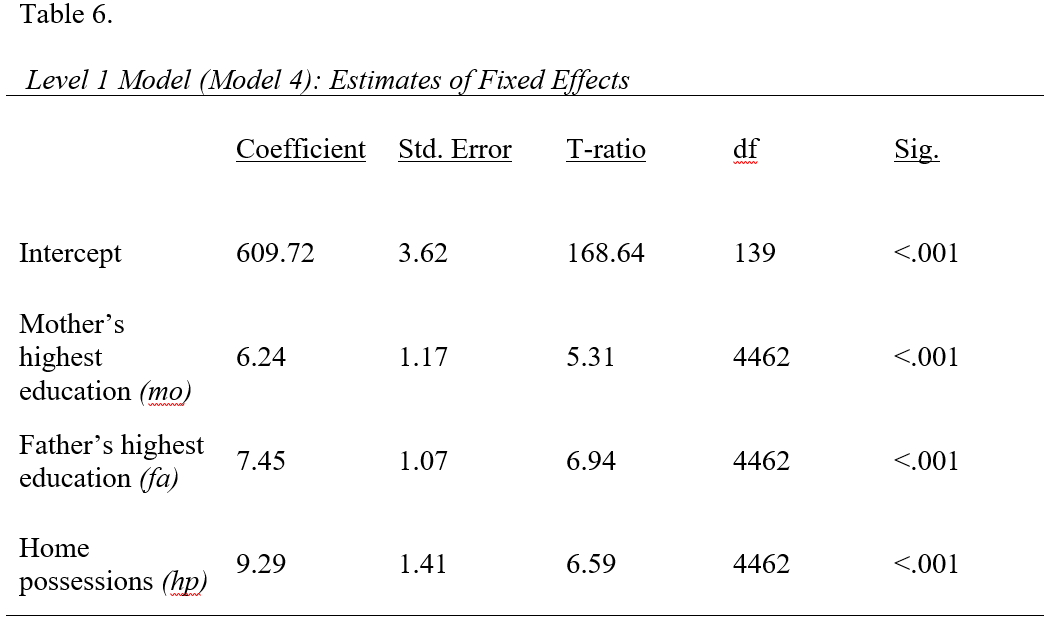}
\end{figure}

\begin{figure}[htbp]
	\centering
	\includegraphics[width=\textwidth]{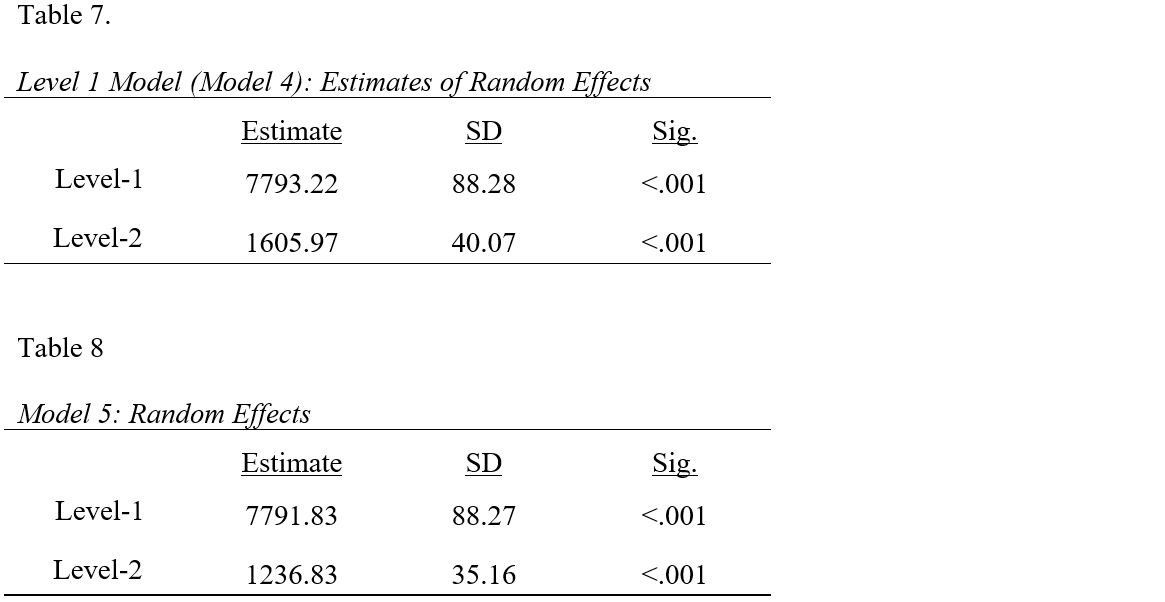}
\end{figure}

\begin{figure}[htbp]
	\centering
	\includegraphics[width=\textwidth]{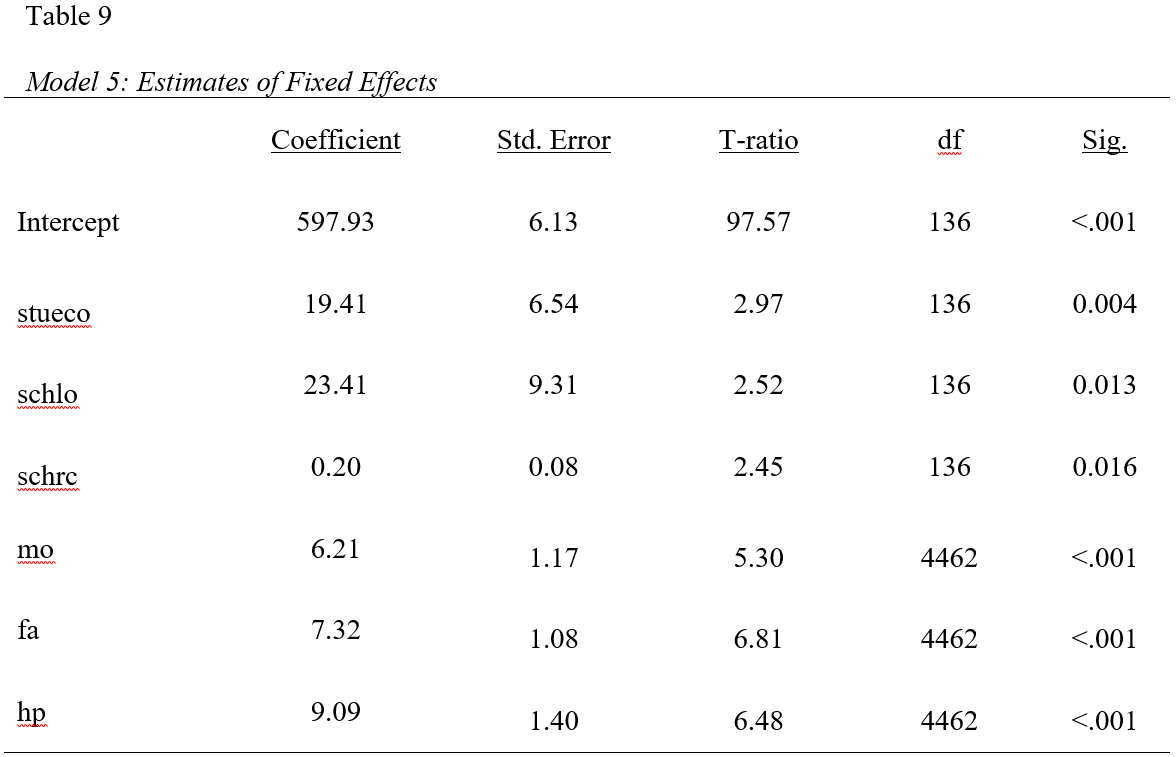}
\end{figure}

\end{document}